\documentclass[manuscript]{aastex}

\begin{document}

\markboth{C. Stubbs}
{Addressing the Crisis in Fundamental Physics}

%
%

\title{ADDRESSING THE CRISIS IN FUNDAMENTAL PHYSICS }

\author{CHRISTOPHER W. STUBBS}

\affil{Department of Physics
\affil{\&}
\affil{Department of Astronomy}
\affil{Harvard University}
\affil{17 Oxford Street, Cambridge MA USA 02138}
\affil{\it cstubbs@fas.harvard.edu}}

\maketitle

\begin{abstract}
The observation that the expansion of the Universe
is proceeding at an ever-increasing rate, the ``Dark Energy''
problem, constitutes a crisis in fundamental physics that is 
as profound as the one that preceded the advent of quantum 
mechanics. Cosmological observations currently favor a Dark Energy equation of 
state parameter $w=P/\rho=-1$. Awkwardly, this is the value that has the least 
ability to discriminate between alternatives for the 
physics that produces the observed accelerating expansion. 
If this result persists we therefore run a very real
risk of stagnation in our attempt to better understand the 
nature of this new physics, unless we uncover another piece of 
the Dark Energy puzzle. I argue that precision fundamental measurements in space 
have an important role in addressing this crisis. 
\end{abstract}

\section{Introduction: The ``Standard Models'' of Particle Physics
and Cosmology }

Over the course of one human lifetime we have developed 
two powerful ``standard models''. 
One pertains to particle physics while
the other is cosmological. On the particle physics front we 
know of 3 families of quarks and leptons, which interact via 
exchange bosons. This picture is known to be incomplete: 
We don't know how to fit gravity into this scenario, and 
the dark matter continues to elude us. 

On the cosmological side, driven in no small measure by the recent WMAP measurements
of the structure in the cosmic microwave background \cite{WMAP}
we now assert that: 

\begin{itemize}

\item{} Our geometrically flat Universe started in a hot big bang 
13.7 Billion years ago.

\item{} The matter component of the Universe is dominated by
dark matter, which is most likely outside the scope of the particles
that make up the standard model of particle physics. 

\item{} Luminous matter comprises only a few percent of the total mass of the Universe. 

\item{} The evolution of the Universe is increasingly dominated 
by the phenomenology of the vacuum. 

\end{itemize}

This consensus cosmology is, however, ludicrous. The assertion 
that two regions of the vacuum experience a mutually repulsive 
gravitational interaction is, well, repulsive. With apologies to 
Kirk, Scotty and especially Mr. Spock, it's like living in 
a bad episode of Star Trek. It is worth asking, therefore, why
this preposterous consensus has emerged, and on the basis
of what experimental data. 

The ingredients of the Universe are most conveniently expressed
in terms of a cosmic sum rule, where
$$\Omega_k + \Omega_\Lambda + \Omega_m =1,$$ 
with $\Omega_k$ representing the contribution from 
any underlying curvature in the underlying geometry of the Universe, 
$\Omega_\Lambda$ reflects the contribution of the Dark Energy
component, and $\Omega_m$ accounts for the matter density
of the Universe (in units of the critical density).

\section{The Observational Evidence for Dark Energy}

The first claims for an accelerating expansion of the Universe 
were made in the late 1990's  by two teams \cite{SN1}$^,$\cite{SN2} 
that used type Ia supernovae to probe the history of cosmic expansion. Supernovae at redshifts around 0.7 were seen to 
have luminosities that were about 20\% fainter than 
expected in a well-behaved Universe. 

As with nearly all results in observational cosmology, the 
supernova measurements have the potential for unappreciated systematic error, and the type Ia Hubble diagram {\it alone} is (in my view) 
insufficient to compel us to believe in the Dark Energy, with 
$\Omega_\Lambda \sim 0.7$.

The sobering fact is that essentially all subsequent cosmological 
data sets drive us to the conclusion that the Universe has recently
entered a time in which the scale factor is growing exponentially. The supernova data, measurements of 
primordial element abundances, determinations of the overall 
mass density of the Universe, and the structure seen in the  microwave background all support this conclusion \cite{RMP}. 

Cognizant of the danger of making a list that appears to be comprehensive but may not be, the possibilities for the 
underlying physics include:

\begin{enumerate}

\item{} A classical cosmological constant, residing in the 
gravitational physics sector, 

\item{}Vacuum energy effects, arising from quantum mechanical 
fluctuations, and

\item{}A modification of gravity that is manifested on the 
cosmological scale. 

\end{enumerate}

The good news aspect of this situation is that we have clear evidence of new physics. The bad news is that we presently have no idea what it means. 

\vskip 2.0in

\subsection{Dark Energy Constitutes a Crisis in Fundamental Physics}

{\it In my opinion the Dark Energy situation constitutes a crisis
in fundamental physics that is every bit as profound as that which 
preceded the advent of quantum mechanics. }

One might wonder
if perhaps we are somehow being misled by the data, or that the data are
being misinterpreted. The convergence of results from such a wide
array of measurement techniques argues against this, but we should nevertheless retain our sense of scientific scepticism. Under such circumstances
we might turn perhaps to theory for some guidance... 

\section{The Present Theoretical Situation}

From the our current perspective there are only
two natural theoretical values for $\Omega_\Lambda$, the apparent
contribution of the vacuum energy in units of the critical 
density. The first of these natural values comes from integrating the effect of all known quantum mechanical fluctuations, up to the 
Planck mass. The resulting value of $\Omega_\Lambda = 10^{120}$
is so preposterous that it's a non-starter. The other natural 
value presupposes that some cancellation mechanism steps in 
and performs a precise cancellation, leaving $\Omega_\Lambda=0$  (to 120 significant figures!). 

In the context of an observed value of $\Omega_\Lambda \sim 0.7$, 
this presents us with a problem. Some scholars have advocated
that anthropic selection effects provide an explanation for this
rather unexpected value, while others retain hope that a 
future theory will provide some natural explanation. 

Today, in the absence of a clear theoretical framework with relevant predictive
power, the Dark Energy problem is best characterized as a 
data and discovery driven endeavor, spiced with interesting
theoretical speculation. 

\section{The Present Observational Situation} 

At the present time observational cosmologists are concentrating 
on measuring the equation of state parameter of the dark energy, 
$w=P/\rho$, the value of which may help us distinguish between 
the different possible mechanisms that might underlie the Dark 
Energy. Current and upcoming cosmological observations 
exploit gravitational lensing, cluster abundances, 
baryon acoustic oscillations, and the supernova Hubble diagram
to improve our understanding of the Dark Energy. These all
have the merit of being undertaken in a regime where the signal 
is non-zero, but have the shared disadvantage of being susceptible 
to the various sources of systematic error that afflict 
astronomical observations. The desire to discriminate between 
different Dark Energy models will push these techniques 
to their limits.  
 
At the time of this writing (mid 2006) we are starting to see preliminary 
determinations of $w=P/\rho$, the equation of state parameter of the Dark Energy, and 
its evolution over cosmic time. One such example is from the 
ESSENCE supernova survey, in which I participate. When
the supernova data are combined with constraints from 
large scale structure\cite {Gajus}, the data favor $w=-1$. Similar results 
have been reported \cite{CFHT}  by the CFHT Legacy Survey team. Unfortunately
this value that has the {\it least} power to discriminate
the origin of the physics that is driving the accelerating expansion.  

\subsection{It Could Get Grim}

Looking ahead, there is a very real possibility that cosmological 
measurements will show that $w=-1$ at all accessible redshifts. 
This would be a grim circumstance, and our attempt to 
discover the nature of the dark energy through cosmological 
observations would completely stall. 

Although science has very recently made huge strides in 
understanding the nature of the reality we inhabit, there are 
certainly ample examples in the past of scientific fields going 
through 
long periods of stagnation. I fear that if we arrive at
$w=-1$, we may be facing such a fate unless we find another
piece in this jigsaw puzzle. 

\section{Precision Fundamental Physics Experiments and the Dark Energy Problem.}  

Many of the speculative ideas that have been put forward to 
explain the dark energy produce observable effects in other
domains. 
We can therefore hope that a vigorous program of precision 
measurements and tests of fundamental physics might 
produce another anomalous result, which when integrated 
with the cosmological acceleration will lead us to a deeper 
understanding. Of course we don't know where this 
anomaly might arise, and this motivates our undertaking
a broad array of experiments, including

\begin{itemize}

\item{} Testing our understanding of gravity on all scales, 
including the inverse square law, the equivalence principle, 
the strong gravity regime, and gravitomagnetism. 

\item{} Direct tests of the fabric of spacetime, including 
geometrical flatness tests, precision clock experiments, time delay measurements, etc. 

\item{} Tests of fundamental symmetries and Lorentz invariance, 

\item{} Probing the nature of vacuum fluctuations and Casimir forces. 

\end{itemize}

Many of these projects have a long history in the precision 
measurement and fundamental physics arenas. In my
opinion the broader community's appreciation of their importance is likely to grow as we collectively turn to face the unexpected crisis posed by the Dark Energy. 

\section{The Role of Space-based Projects}

We should distinguish at the outset between space-based 
{\it observations} and {\it experiments}. On the observational 
side, we exploit the low IR background,  
diffraction-limited imaging, and predictable absence of weather to 
collect photons from distant sources. We then use these observations
to infer the properties of the Universe we inhabit. 

This is to be contrasted with space-based {\it experiments}, where
the absence of seismic perturbations, 
the microgravity environment, and 
access to various solar system gravitational potentials are 
combined with precision apparatus to pose well-defined 
experimental questions of nature. 

The ability to carry out precision measurements, in which we
can explicitly test the system's susceptibility to potential 
sources of systematic error, gives experiments 
a special role to play in probing for a deeper understanding
of Dark Energy. 

Our challenge is to identify those instances where we can 
realize large gains by combining the tools and techniques from 
the precision metrology and fundamental measurement 
communities with the special environmental aspects of space. 
Once these opportunities are identified we face our next 
daunting challenge, namely prioritizing the different options 
when we (for now, at least) have no idea where the 
signature for new physics might next emerge. As articulated
by Dr. Marburger in his opening remarks at this meeting, 
far better that we scientists make that assessment than 
leaving it to lobbyists and legislators. 

\section{An Example-- Laser Ranging in the Solar System.}

This meeting has numerous interesting presentations on 
existing and proposed fundamental physics projects, 
many of which might exhibit the next piece of evidence for
new physics. Let me pick one illustrative example and 
indicate how it links to the Dark Energy problem. 

Smith et al. have recently demonstrated\cite{lasers} the successful 
exchange of laser communication signals with a spacecraft
a distance 24 million km from Earth. See also the talk by 
Degnan in these proceedings. This gives reality to the notion of ``piggybacking'' a precision laser ranging capability on 
laser communication links. If NASA holds to its current 
goals of missions to the moon and Mars, we could imagine 
adding a fundamental physics aspect to these flight opportunities. 
(See the paper by Merkowitz in this volume, and references
therein).  

The scientific merit of an aggressive and coordinated solar system  
ranging campaign has been outlined\cite{Ken} by Nordtvedt, using the
various gravitational interactions between the bodies in the solar system.  
Performing a global fit to ranging data between the
Earth, the moon, and the other planets  
will allow us to perform a comprehensive test of the basic 
foundations of gravity. 

Successfully undertaking projects of this sort will require a coordination of efforts individuals and teams drawn from diverse communities, including: 

\begin {itemize}

\item{} Precision measurement and metrology,

\item{} Atomic-Molecular-Optical physics,

\item{} Astronomy,

\item{} Gravitational theory and numerical analysis,

\item{} Particle physics,

\item{} Precision engineering, and  

\item{}Astronautics and spacecraft engineering.

\end {itemize}

This meeting is a great opportunity for us to renew old partnerships, and to build new ones. 

\section{An Exhortation}

Let me then end with an exhortation. We are in the midst of a 
profound crisis in fundamental physics. On the theoretical side, 
we know that our two primary triumphs, namely general relativity and 
quantum mechanics, do not work well together. On the 
observational side we have the challenge of 
the increasing evidence for the 
accelerating expansion of the Universe. I think there is a 
very real possibility that the observational cosmologists will, 
in the decades ahead, present us with increasingly 
precise measurements of $w=-1$, with no evidence for
evolution over cosmic time. 

This will require that we push the frontiers of fundamental 
physics in order to search for the next piece of the 
Dark Energy puzzle. In my view the task of this meeting
is to identify and refine the concepts that stand the best
chance of capitalizing on space flight opportunities to 
address this crisis in fundamental physics. 

\section{Acknowledgments}

 I would like to thank the organizers for putting together such 
 a stimulating meeting. In addition I would like to thank
 both Slava Turyshev and Thomas Murphy for their invitation, 
 and for their generous help, encouragement, and 
 valuable comments while I was preparing this talk. 
I am also grateful to Harvard University, to the Department of 
Energy (through their grant to Harvard's Laboratory for Particle 
Physics and Cosmology) and to the National Science 
Foundation (under grant AST-0507475) for supporting 
my own work on the Dark Energy problem.

\end{document}